\documentclass[journal]{IEEEtran}
\usepackage{mathrsfs}
\usepackage{amssymb}
\usepackage[mathcal]{euscript}
\usepackage{cite}
\usepackage{graphicx}
\usepackage{psfrag}
\usepackage{subfigure}
\usepackage{url}
\usepackage{booktabs}
\usepackage{stfloats}
\usepackage{amsmath}
\usepackage{float}
\usepackage{array}
\usepackage{amsthm}
\usepackage{booktabs}  
\usepackage{algorithm} 
\usepackage{setspace}
\usepackage{algorithmic} 
\usepackage{amsmath,amsfonts,amssymb,amsbsy,bm,paralist,theorem,ifthen,color}
\usepackage{color}
\usepackage{graphicx}
\usepackage{color}
\usepackage{placeins}
\usepackage{float}
\usepackage{hyperref}
\usepackage{tabularx,colortbl}
\usepackage{multirow}
\usepackage{multicol}
\usepackage{arydshln}
\usepackage{makecell}
\usepackage{stfloats}
\usepackage[table,xcdraw]{xcolor}

\title{Electromagnetic Information Theory: \\Fundamentals and Applications for 6G Wireless Communication Systems}
\author{Cheng-Xiang~Wang,~\IEEEmembership{Fellow,~IEEE,}
	Yue Yang,
	Jie Huang,~\IEEEmembership{Member,~IEEE,}
	Xiqi Gao,~\IEEEmembership{Fellow,~IEEE,}\\
	Tie Jun Cui,~\IEEEmembership{Fellow,~IEEE,}
	and Lajos Hanzo,~\IEEEmembership{Life Fellow,~IEEE}
        
\thanks{
	
	C.-X. Wang (corresponding author), Y. Yang, J. Huang, and X. Q. Gao are with the National Mobile Communications Research Laboratory, School of Information Science and Engineering, Southeast University, Nanjing 210096, China, and also with the Purple Mountain Laboratories, Nanjing 211111, China (email: \{chxwang, yueyang, j\_huang, xqgao\}@seu.edu.cn).
	
	T. J. Cui is with the State Key Laboratory of Millimeter Wave, School of Information Science and Engineering, Southeast University, Nanjing 210096, China (e-mail: tjcui@seu.edu.cn).

    L. Hanzo is with the School of Electronics and Computer Science, University of Southampton, Southampton, U.K. (e-mail: lh@ecs.soton.ac.uk).

}}

\begin{document}

\maketitle

\begin{abstract}
In wireless communications, electromagnetic theory and information theory constitute a pair of fundamental theories, bridged by antenna theory and wireless propagation channel modeling theory. Up to the fifth generation (5G) wireless communication networks, these four theories have been developing relatively independently. However, in sixth generation (6G) space-air-ground-sea wireless communication networks, seamless coverage is expected in the three-dimensional (3D) space, potentially necessitating the acquisition of channel state information (CSI) and channel capacity calculation at anywhere and any time. Additionally, the key 6G technologies such as ultra-massive multiple-input multiple-output (MIMO) and holographic MIMO achieves intricate interaction of the antennas and wireless propagation environments, which necessitates the joint modeling of antennas and wireless propagation channels. To address the challenges in 6G, the integration of the above four theories becomes inevitable, leading to the concept of the so-called electromagnetic information theory (EIT). In this article, a suite of 6G key technologies is highlighted. Then, the concepts and relationships of the four theories are unveiled. Finally, the necessity and benefits of integrating them into the EIT are revealed.
\end{abstract}


\section{Introduction}
The sixth generation (6G) technologies are in the preliminary stages of exploration. The research, development, and standardization of 6G technologies require major theoretical breakthroughs. Electromagnetic theory and information theory constitute a pair of pivotal  theories in wireless communications. They are intricately interlinked by wireless propagation channel modeling theory and antenna theory. Electromagnetic theory describes the generation and propagation of electromagnetic waves, while information theory quantifies the amount of information transmitted. Wireless propagation channel modeling theory characterizes the wireless propagation channels between the transmitter (Tx) and receiver (Rx), excluding the Tx and Rx antennas. As a further pivotal component, antenna theory considers the design, analysis, and characterization of antennas. Finally, the wireless propagation channel links the Tx and Rx antennas and carries information via electromagnetic waves, acting as the bridge connecting the four theories. The integration of the four theories is termed as electromagnetic information theory (EIT), which was first proposed in \cite{ref_MD} and \cite{ref_MD2} with the objective of analyzing the constraints and limitations of information transmission caused by electromagnetism. Then, harnessing hitherto unexploited dimensions of electromagnetic waves, such as orbital angular momentum (OAM), constitutes a compelling application of EIT. Furthermore, the employment of metamaterials relying on the marriage of physical and mathematical principles promotes the further development of EIT \cite{ref_RIS}. The EIT-based system model for multiple-input multiple-output (MIMO) systems may, however, fail to obey some conventional assumptions of wireless communications\cite{ref_SW}. In summary, EIT attempts to model general channels in a comprehensive manner. Some recent advances and open problems of EIT were shown in \cite{ref_EIT}.

Up to the fifth generation (5G) wireless communication networks, these four theories have been developing relatively independently. However, 6G aims for supporting the intelligent Internet of everything. Furthermore, 6G  will fully utilize all the frequency bands, spanning from sub-6 GHz to optical wireless bands. In terms of coverage, 6G will be extended from local terrestrial coverage to space-air-ground-sea global coverage. Furthermore, 6G will intrinsically amalgamate technologies such as integrated sensing and communication (ISAC), ultra-massive MIMO, holographic MIMO schemes, and reconfigurable intelligent surfaces (RISs), aiming for the provision in support of intelligent and secure networks. In short, the 6G visions can be summarized as global coverage, all spectra, full applications, and strong security \cite{ref_6Gvision}. Correspondingly, 6G channels can be summarized as all frequency band channels, global coverage scenario channels, and full application scenario channels \cite{ref_6GPCM}.

In 6G space-air-ground-sea wireless communication networks, the base stations (BSs) may be mobile in low-earth orbit (LEO) satellite and vehicular communications. Hence, the underlying wireless channels tend to be of three-dimensional (3D) mature in space, potentially necessitating the acquisition of the channel state information (CSI) and channel capacity calculation at any location in the 3D space. The underlying electromagnetic theory relies on continuous time and space, facilitating the calculation of the continuous-space electromagnetic field distribution. However, relying solely on electromagnetic theory does not allow us to calculate the continuous-space channel capacity. To achieve their ambitious objective, electromagnetic theory and information theory have to be integrated in 6G to solve the above problems. Furthermore, ultra-massive MIMO schemes will further increase the number and size of antennas, while holographic MIMO will reduce the antenna spacing, resulting in tighter integration of antennas and the underlying propagation environment. However, the integration of these four theories poses challenges, while creating some new opportunities for joint antenna and system design. In summary, 6G space-air-ground-sea networks exhibit extreme technical requirements, which are beyond the scope of each of the individual theories. It is necessary to integrate the four theories to form the fundamental architecture of EIT.

In a nutshell, we illustrate the fundamentals, requirements, and applications of EIT. Firstly, a range of 6G techniques, including holographic MIMO, ISAC, RIS, space-air-ground-sea integrated networks, and experimental channel measurement verifications are illustrated. Secondly, the relationships of the four fundamental theories and their gaps are investigated. Finally, the compelling case for integrating the four theories into EIT is discussed, and the potential performance benefits conclude our discourse.

\section{6G Key Technologies and Challenges}
	In this section, 6G key technologies and challenges are highlighted, including holographic MIMO, ISAC, RIS, space-air-ground-sea integrated networks, and also experimental channel measurement verifications. 
	
	\subsection{Holographic MIMO Aided Communications}
	Holographic MIMO schemes evolved from traditional massive MIMO, which rely on an infinite number of antenna elements in a compact space to realize a continuous-aperture array. Therefore, classic discrete channel modeling is not applicable, it has to be integrated with continuous-space electromagnetic theory. When the antenna elements are displaced closer and the number is increased, the Rayleigh distance becomes higher. The mutual coupling and near-field effects become more obvious. However, the influence of mutual coupling and near-field effects on the communication performance is not clear from the perspective of antenna design and channel modeling. For example, in terms of antenna design, the mutual coupling effect has to be mitigated as much as possible, because it can degrade the performance. However, as for channel modeling, recent research concluded that mutual coupling may improve the channel capacity of holographic MIMO, in contrast to conventional antenna design. Therefore, constructing a unified model for mutual coupling from the dual perspective of antennas and channels is important for holographic MIMO channel modeling. Additionally, precoding schemes considering mutual coupling also require further study for system performance enhancement.
	
	\subsection{ISAC}
	ISAC is capable of improving spectral efficiency, hardware efficiency, and information processing efficiency in conjunction with sophisticated spectrum sensing. However, it requires accurate 3D CSI in real time for positioning, ranging, velocity estimation, imaging, detection, recognition, and environmental reconstruction. As a benefit, it provides situational awareness in the physical world. Therefore, the relationships between the physical environment and the statistical properties of the wireless channel model are critical for high-precision CSI acquisition. ISAC requires more detailed analysis and the extraction of information from electromagnetic waves to effectively measure the information contained in the continuous space. Due to the heterogeneous characteristics of ISAC, the electromagnetic waves contain a wealth of information such as time, space, structure, and material of the environment encountered. Traditional communication and sensing schemes are not applicable, and it is necessary to establish the unique information theory of ISAC systems based on electromagnetic theory to characterize the communication and sensing performance of the system.
	\subsection{RIS Channel Modeling}
	As an extension of multi-antenna technology, RISs have attracted wide attention \cite{ref_RIS}. They are capable of achieving coverage expansion, including non-line-of-sight scenario enhancement and supporting edge users. They also support high-precision positioning. RISs constitute a multidisciplinary technology, including material science, electromagnetism, information-, electronics-, and communication-engineering, requiring substantial further research. However, there are still numerous challenges involving their theoretical models and applications. RISs are composed of a large number of passive reflection units, and the reflection coefficient of each unit has to be adjusted. Moreover, the RIS channel is of full rank, which can increase the channel capacity. In addition, RIS channel measurements only focus on channel large-scale fading, with less attention dedicated to small-scale channel fading, and do not study the impact of propagation environment and important antenna parameters on the statistical properties of the RIS channel model.
	\subsection{Space-Air-Ground-Sea Integrated Networks}
	The 6G networks are three-dimensional space-air-ground-sea integrated networks. The BSs constituted by satellites and unmanned aerial vehicles tend to move continuously. Therefore, space-air-ground-sea integrated networks have to consider non-stationarity in the space-, time-, and frequency-domains, while relying on the intelligent scheduling of network resources. Space-air-ground-sea integrated networks have mobile base stations and users, and they can be distributed everywhere. Therefore, it is important to obtain CSI in continuous space. It also has to ensure high-reliability and continuous-space multi-user communications, so the network reliability and stability are important challenges. Their security issues including their secrecy capacity also necessitates a community-wide effort.    
	\subsection{Experimental Channel Measurement Verifications}
	The aforementioned key technologies can be verified through 6G wireless channel measurements, for example, holographic MIMO, ISAC, and RIS channel measurements. Firstly, holographic MIMO technology can sense complex electromagnetic environments and realize the real-time predictive analysis of electromagnetic space. Holographic MIMO channel measurements can infer the transmission rate of continuous-space channels having extremely high spatial resolution and support continuous-space channel modeling and antenna design. Secondly, by harnessing ultra-massive MIMO antenna arrays, ISAC channels can be characterized. Thirdly, through RIS wireless channel measurements, the mechanism of how the physical environment affects the electromagnetic waves can be explored. In RIS-aided systems, different numbers, positions, and topologies of RIS units may be used at the Tx and Rx, where the electromagnetic wave propagation is manipulated by reflected phase control. Therefore, RIS channel measurements can support the expanded applications of Maxwell's equations. Finally, the influence of important antenna parameters such as their mutual coupling coefficient, radiation efficiency, and bandwidth on the radio channel's statistical properties and channel capacity can be explored. 

\section{Four Fundamental Theories}
In this section, four fundamental theories are illustrated and their relationships with key technologies are investigated.

\subsection{Electromagnetic Theory}
Electromagnetic theory characterizes the generation and propagation of electromagnetic waves and studies the relationships between various physical quantities in electromagnetic fields. It hinges on Maxwell's equations partial differential equations and integral equations describing the relationship of electric and magnetic fields. Electromagnetic theory can be applied to deterministic ray tracing propagation channel modeling, electromagnetic field analysis, as well as to circuit and chip design. In 6G communication systems, both ISAC and holographic MIMO schemes rely on continuous-space electromagnetic fields and CSI of the environments. However, applying electromagnetic theory individually to obtain the continuous-space electromagnetic fields and CSI would result in excessive computational complexity.
\subsection{Information Theory}
Information theory deals with the transmission and processing of information. It focuses on the efficient and accurate transmission of information under the constraints of limited bandwidth and power. The classical Shannon capacity quantifies the maximum error-free transmission rate within a constrained bandwidth and power in additive white Gaussian noise (AWGN) channel\cite{ref_Shannon}. Information theory characterized the channel capacity of a whole raft of wide-sense stationary (WSS) fading channels of both single- and multi-user systems, as well as their degree of freedom (DoF). However, 6G faces the hitherto unprecedented requirements of ultra-massive MIMO, high-mobility, millimeter-wave/terahertz communications, and space-air-ground-sea integrated networks, relying on ultra-massive MIMO schemes exhibiting space-time-frequency non-stationarities. Furthermore, the BSs and users are roaming continuously in LEO satellite and vehicular communications in the 3D space. The near-field and non-stationary channel capacities cannot be calculated by using information theory alone, and there is no unified formula for multi-user/network capacity calculation in the continuous space.
\subsection{Wireless Propagation Channel Modeling Theory}
Wireless propagation channel models describe the channels between the Tx and Rx antennas, but exclude the Tx and Rx antennas. Channel fading can be classified as large-scale fading - including the path loss and shadow fading - and small-scale fading. Large-scale fading is utilized in wireless network planning and optimization. Small-scale fading is important in the context of channel estimation, modulation, and coding. Wireless propagation channel models include deterministic channel models, geometry-based stochastic models (GBSMs), correlation-based stochastic models (CBSMs), and beam domain channel models (BDCMs). These channel models also consider a part of the antenna radiation parameters. New antenna design paradigms such as holographic MIMO and ultra-massive MIMO are closely related to the communication environments. However, applying classic wireless propagation channel modeling theory in isolation cannot consider all antenna parameters and their influences on the associated statistical properties, hence the associated channel capacities are unknown.

\begin{figure*}[tb]
	\centerline{\includegraphics[width=6.7in]{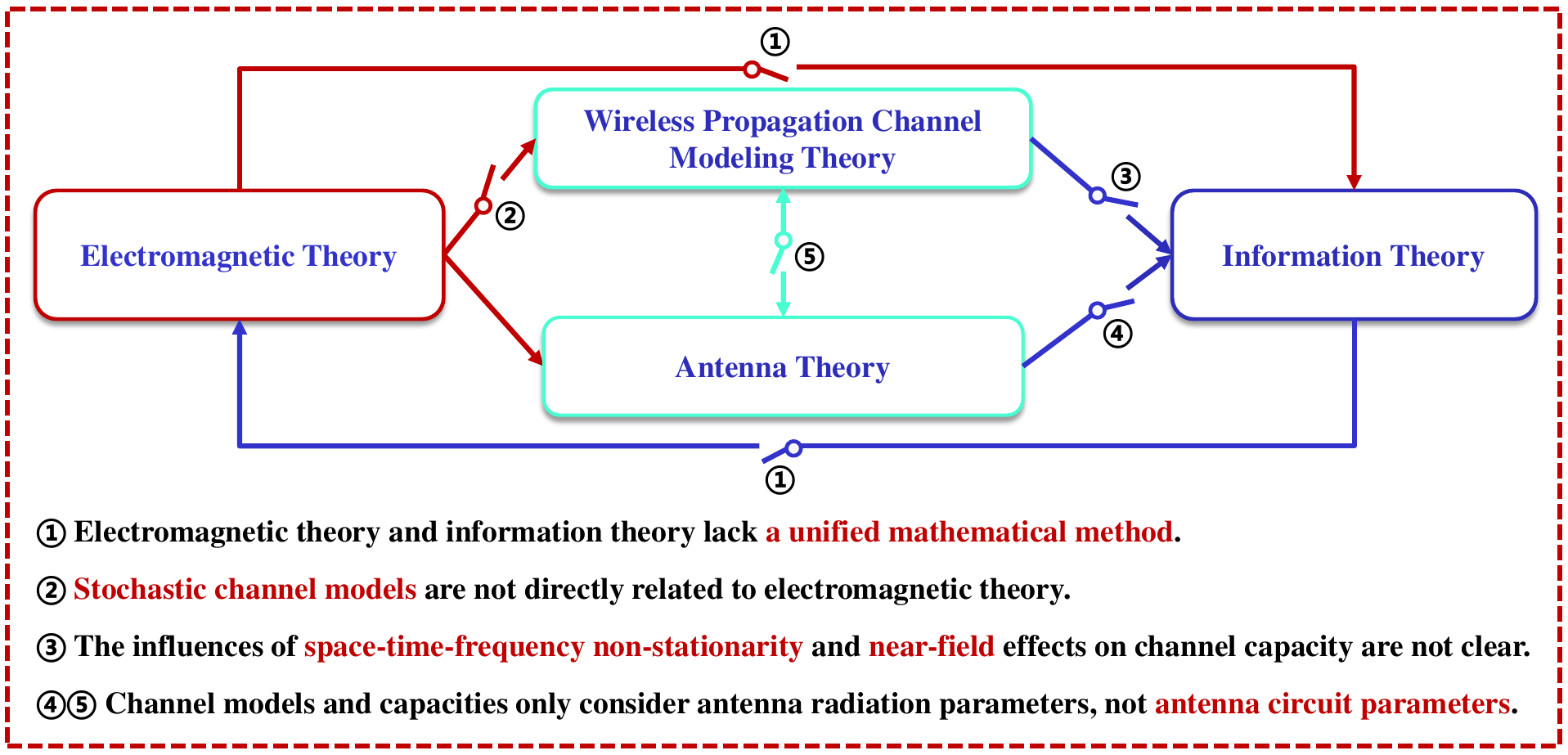}}
	\caption{Relationships and gaps between electromagnetic theory, information theory, wireless propagation channel modeling theory, and antenna theory.}
	\label{1}
\end{figure*}

\begin{table*}[t]
	\centering
	\caption{The 6G technical challenges and limitations of a single theory.}
	\renewcommand\arraystretch{1.15} 
	\begin{tabular}{|m{5.5cm}|m{6.3cm}|l|}
		\hline
		\textbf{6G Technical Challenges}\centering&\textbf{Limitations of A Single Theory}\centering
		&\textbf{\ \ Directions of Theory Integration}
		\\ \hline
		
		Both ISAC and holographic MIMO schemes require continuous-space electromagnetic fields and CSI of the environments.
		&Electromagnetic theory applies to deterministic channel modeling and leads to high computational complexity. &\begin{tabular}[l]{@{}l@{}}$\bullet$ The expanded applications of Max-\\well's equations \\ $\bullet$ Continuous-space electromagnetic \\channel models\end{tabular} 
		\\ \hline
		
		\begin{tabular}[l]{@{}l@{}}$\bullet$ Ultra-massive MIMO, high mobility, \\ and wideband communications lead to\\channel non-stationarities.\\$\bullet$ Multi-user/network capacity has to \\be considered in space-air-ground-sea \\integrated networks.\end{tabular} &\begin{tabular}[l]{@{}l@{}}$\bullet$ Information theoretical calculations of the \\near-field and non-stationary channel capaci-\\ties become excessively complex.\\$\bullet$ Information theory lacks a unified formula\\ for multi-user/network capacity calculation.\end{tabular}&\begin{tabular}[l]{@{}l@{}}$\bullet$ Non-stationary channel capacity\\ calculation \\$\bullet$ Continuous-space electromagnetic\\ channel capacity analysis\end{tabular}
		\\ \hline
		
		Upon increasing the number of elements in ultra-massive MIMO and holographic MIMO, the antennas become intricately combined with the environment. & Wireless propagation channel modeling only considers the antenna radiation parameters, but not the antenna circuit parameters. & Radio channel models
		\\ \hline
		
		Design of new antennas such as ultra-massive MIMO and holographic MIMO antennas has to consider the features of the Tx and Rx and channel statistics.& Traditional antenna theory only considers features of the single-sided Tx or Rx and ignores the influence of antenna circuit parameters such as resonant frequency, impedance, and bandwidth on channel capacity. & \begin{tabular}[l]{@{}l@{}}Wide-sense antenna theory\end{tabular}
		\\ \hline 	
	\end{tabular}
	\label{tab1}
\end{table*}

\subsection{Antenna Theory}
Antenna theory deals with the design, analysis, and characterization of antennas. The main design techniques rely on radiation field analysis with the objective of meeting the requirements of communication systems. Antennas generate and receive electromagnetic waves, linking the Tx and Rx with the aid of specifically designed transmit and receive beamforming pattern. In 6G communication systems, novel technologies such as RIS, ultra-massive MIMO, and holographic MIMO have emerged and they have to consider the features of both the Tx and Rx antennas as well as channel statistics. However, antenna theory only considers single-sided Tx or Rx antenna design and ignores the influence of all antenna parameters on the channel capacity.

\subsection{Relationships and Gaps Among the Four Theories}
Compared to 5G communication systems, 6G faces radical requirements in support of ISAC, ultra-massive MIMO, holographic MIMO, RIS, and space-air-ground-sea integrated networks. These demanding applications rely on an amalgam of individual theories. Their relationships and gaps are highlighted at a glance in Fig.~\ref{1}. Electromagnetic theory is based on mathematical calculus, while information theory relies on probability theory and linear algebra theory. However, they lack a unified mathematical apparatus. The classic ray tracing based wireless propagation channel modeling relies on a simplified form of electromagnetic theory in the high-frequency band. However, GBSMs, CBSMs, and BDCMs are not directly related to electromagnetic theory. Naturally, the specific channel model will directly affect the capacity, but the influence of space-time-frequency non-stationarity and near-field effects on channel capacity has not been quantified. In addition, the antenna circuit parameters, such as the impedance and resonant frequency are not included in the wireless propagation channel models and capacity expressions. Again, since the above four theories were developed independently, we set out to intrinsically amalgamate them for 6G wireless communication systems.

\section{EIT: Integration of the Four Theories}
In this section, the necessity and methods to integrate the four theories will be investigated. The technical 6G challenges and directions of theory integration are summarized in Table~\ref{tab1} at a glance. The benefits of integration can be outlined as twinning electromagnetic theory and information theory, increasing the channel capacity, accomplishing wireless channel modeling theory, and facilitating antenna design. Therefore, the benefits of EIT will guide antenna design and enhance the performance of 6G wireless communication systems. For example, the performance of channel estimation can be improved when integrating EIT into classical minimum mean square error (MMSE) channel estimation.

\subsection{Twinning Electromagnetic Theory and Information Theory}
ISAC and holographic MIMO schemes have to obtain continuous-space CSI and channel capacity in support of 6G communication systems. Classic electromagnetic theory is based on continuous time and space. Maxwell's equations are the most fundamental equations of classic electromagnetic theory. They are mainly applied to macroscopic scale fields. When the scale is reduced to a microscopic scale, the local fluctuations in fields become non-negligible. In high-mobility, millimeter wave-aided wideband scenarios, it is necessary to consider the effects of electromagnetic coupling and polarization, the random time-varying characteristics, and time-frequency non-stationarities. Classic Shannon’s information theory only considered simple channel, such as AWGN channels, and assumed wide-sense stationary signals. Since some key 6G technologies, such as ISAC, RIS, massive MIMO, and holographic MIMO, need to consider more realistic environments and non-stationary channels, classic information theory cannot accurately represent the capacities under these circumstances and need to change in form. Furthermore, wireless communication systems designed by classical information theory do not give full cognizance to the specific nature of electromagnetic waves, complex propagation characteristics, and mechanisms brought about by electromagnetic field effects. By integrating these two fundamental theories, information theory can be expanded from WSS channels having discrete distributions to non-stationary channels associated with continuous distributions. A continuous-space electromagnetic channel model describing the relationship between the radiated field and arbitrary source current distribution by dyadic Green's function was proposed in \cite{ref_jensen}. It considered the constraints of basis functions and the average radiated power and maximized the channel capacity by adopting optimal antenna design. The discrete-space single-user information theory can be evolved into continuous-space multi-user/network information theory.

Electromagnetic theory can be expanded from static to high-mobility media. By calculating the amount of mutual information between incident waves, reflected waves, transmitted waves, and scattered waves, we can study the influence of the spatial distribution of scatterers. Then, we can study the application of Maxwell's equations to RIS and other new technologies. The potential performance and capacity limits may be accurately described. In \cite{ref_RIS}, the authors proposed the concept of information metamaterials and tried to integrate electromagnetic theory and information theory. By constructing a digital space on the physical space of metamaterials, the electromagnetic waves can be beneficially manipulated for increasing the spectral efficiency and for reducing the system costs. Information metamaterials may be viewed as an expanded application of Maxwell's equations, potentially redefining of the architecture of information systems. The characterization and processing of wireless communication signals is often based on one-dimensional vector assumptions and cannot accurately characterize electromagnetic fields. Therefore, we intend to characterize the interaction between electromagnetic waves by using signal processing methods to explore the electromagnetic waves through RIS. Then, the end-to-end electromagnetic information transmission mechanism of the communication system is realized.
\subsection{Increasing the Channel Capacity}
6G systems have to support a large number of users and base stations in ubiquitous connection scenarios. Channel capacity is determined by the bandwidth, the number of antennas, the signal-to-noise ratio (SNR), and the channel matrix. When the bandwidth, the number of Tx and Rx antennas, and the SNR are fixed, the method to realize an increased capacity is to appropriately adjust the parameters in the channel matrix $\mathbf{H}$. By changing the parameters of multi-path components, the delay, angle, and power of multi-path between the Tx and Rx antennas are impacted. Thus, the characteristics of the multi-path, the channel correlation, and the channel rank can be adjusted appropriately to achieve higher capacity.
Therefore, beneficially ameliorating the channel matrix may bring about new gains by increasing the rank and dimension of the channel matrix. 

\begin{figure}[tb]
	\centerline{\includegraphics[width=3.5in]{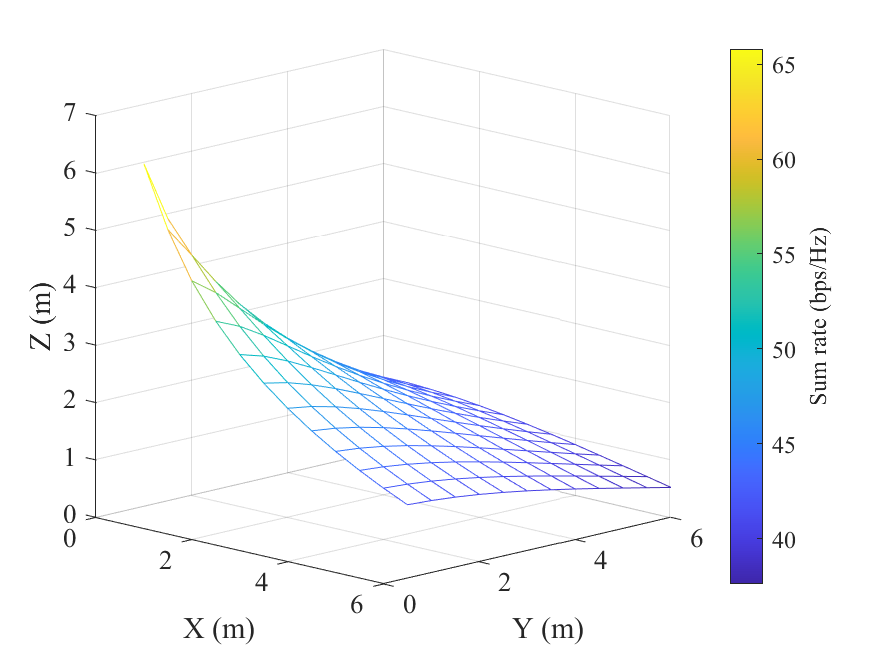}}
	\caption{Capacity of the continuous-space electromagnetic channel.}
	\label{2}
\end{figure}

Holographic MIMO and ISAC schemes have to obtain CSI at any 3D position and transceiver antennas tend to be distributed continuously in holographic MIMO channels \cite{ref_HMIMO}. In the legacy systems, the channel is usually characterized based on the scalar electromagnetic field hypothesis and ignores the vectorial electromagnetic field characteristics of near-field conditions, which limits the DoF of the electromagnetic channel to a certain extent and hence cannot accurately characterize the capacity of continuous-space channels. Therefore, we can calculate the capacity of continuous-space channels based on multi-user/network information theory and establish a unified representation of EIT in the continuous space. For instance, the channel capacity of the continuous-space electromagnetic channel at SNR = 10~dB is shown in Fig.~\ref{2}. Continuous-space electromagnetic channel models are capable of expanding the dimension of wireless channels and hence increase the channel capacity.

More explicitly, both the amplitude and phase of electromagnetic waves may be manipulated by the circuit response to increase the rank of the channel matrix. For example, both RIS and OAM are capable of constructing a full-rank matrix, and the resultant capacity gain can be exploited. Furthermore, channel capacity gains may also be gleaned by using the multi-polarized antennas and by increasing the number of electromagnetic wave transmission modes. With the developments of ultra-massive MIMO and holographic MIMO schemes, the near-field spherical wavefront and mutual coupling effects are significant. The introduction of a mutual coupling matrix may reduce the correlation between channels and hence increase the channel capacity. When considering the near-field steering vectors of the antenna array, the channel may contain more rich angle-of-arrival information than that in the far field, so the channel capacity can also be increased. 

In traditional information theory, there are bottlenecks in the capacity calculation of non-stationary space-time-frequency channels. In near-field communications, the non-stationarity effect is more obvious. However, the non-stationary channel capacity of near-field conditions is unknown, but it is crucial for the overall design of 6G systems. To fill this knowledge gap, the water-filling algorithm of the space-time-frequency domain and the non-stationary channel capacity has to be studied. Shannon's channel capacity theorem is applicable to WSS channels, but it cannot be applied to non-stationary channels. The family of non-stationary channels can be divided into several WSS sub-channels based on their stationary interval, within which the channel is WSS. Then, the non-stationary channel capacity may be defined as the sum of the WSS sub-channel capacities.

Our study shows that the channel capacity increases, when the number of WSS sub-channels increases, and it reaches its maximum, as the number of WSS sub-channels equals the number of stationary intervals. This illustrates that the non-stationary channel capacity is higher than the conventional channel capacity, which means that the conventional channel capacity calculation underestimates the channel capacity in non-stationary channels. Additionally, the influences of the antenna mutual coupling effect and antenna array topologies on channel capacity are important for non-stationary 6G channels. 

\begin{figure}[tb]
	\centerline{\includegraphics[width=3.5in]{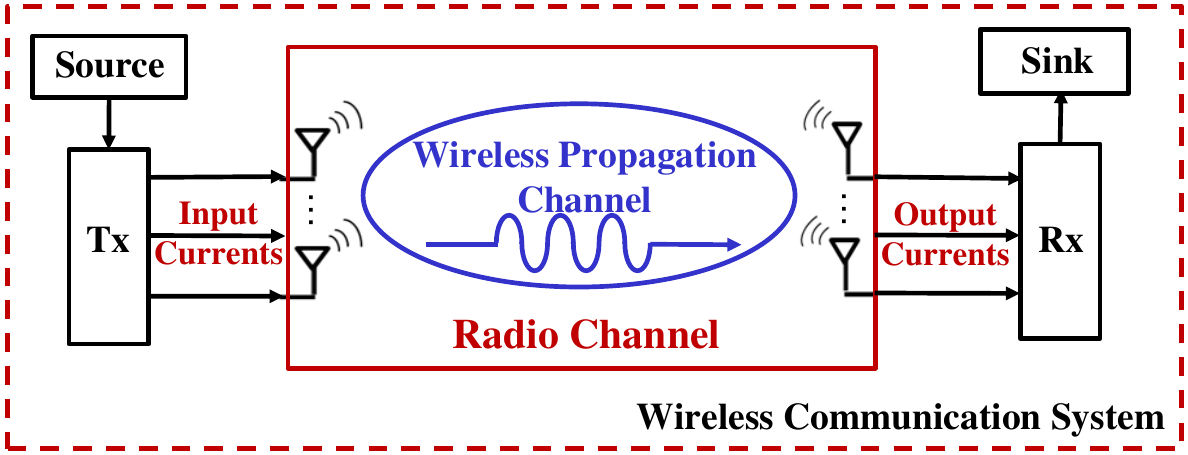}}
	\caption{The description of the wireless propagation channel and radio channel.}
	\label{3}
\end{figure}

\subsection{Accomplishing Wireless Channel Modeling Theory}
Wireless propagation channel modeling considers the basic antenna radiation parameters such as the number of antennas, array type, aperture, and pattern, but it does not consider important antenna circuit parameters. However, wireless propagation channels are intrinsically interlinked with the transceiver antenna array in ultra-massive MIMO channels, hence the impact of antenna parameters on the wireless propagation channels cannot be ignored. The holographic MIMO adopts a large continuous-aperture array, which is closely related to the continuous-space electromagnetic fields. Therefore, we urgently have to construct the relevant radio channel modeling theory combining antenna theory and wireless propagation channel modeling. A birds-eye perspective of the wireless propagation channel and radio channel is shown in Fig.~\ref{3}. The differences between the wireless propagation channel and radio channel are as follows. The wireless propagation channel is that between the Tx and Rx antennas, while the radio channel includes the Tx and Rx antennas and the wireless propagation channel. The integration of wireless propagation channel modeling and antenna theory beneficially augments wireless channel modeling theory.

In \cite{ref_circuit}, a multi-port model based on circuit theory was established for wireless communication systems and the continuous-space electromagnetic channel capacity bound was analyzed. In \cite{ref_chu} and \cite{ref_antenna}, the authors considered the physical limitations of antenna size using the Chu limit to analyze the maximum achievable rate of single-input single-output (SISO) systems. However, only the large-scale fading and the line-of-sight path were modeled. The analysis of the circuit-based MIMO radio channel models remains limited. Therefore, the radio channel models of Fig.~\ref{3} must subsume the wireless propagation channels and the transceiver antennas, describe the circuit-based radio channels and obtain an end-to-end response. Additionally, the radio channel models also have to be explored in the context of more general antenna equivalent circuits and antenna array topologies.

\begin{figure}[tb]
	\centerline{\includegraphics[width=3.5in]{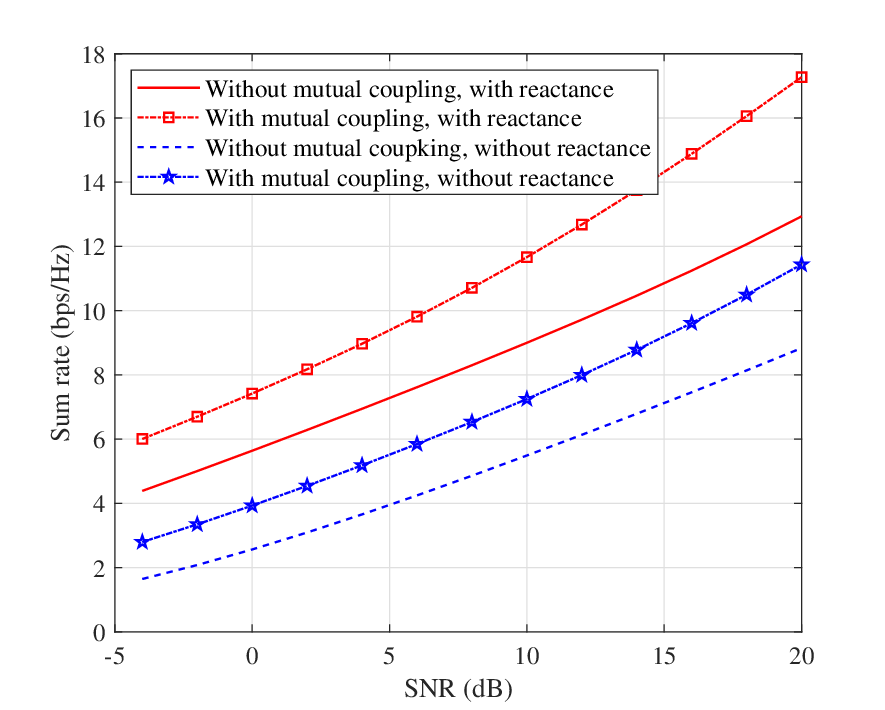}}
	\caption{The impact of mutual coupling and reactance on channel capacity of the radio channel model.}
	\label{4}
\end{figure}

The channel capacity of the radio channel model has to consider the influence of antenna parameters. Therefore, a novel channel capacity formula - including the influence of the transceiver antennas - has to be derived, and the statistical properties of different channels must be studied\cite{ref_radio}. The capacities of the radio channel model considering the antenna circuit parameters, such as the mutual coupling and reactance are shown in Fig.~\ref{4}. Observe that when SNR = 20 dB, the radio channel model considering the reactance increased by about 40\% when neglecting mutual coupling. The influence of mutual coupling and antenna spacing on the channel capacity of the radio channel model is also illustrated. It indicates that the channel capacity may be increased by about 30\%, when considering mutual coupling. These illustrate that the antenna parameters of radio channel models have substantial impacts on channel capacity. Therefore, it is necessary to consider the antenna parameters in channel models.

\subsection{Facilitating Antenna Design}
\begin{figure}[tb]
	\centerline{\includegraphics[width=3.5in]{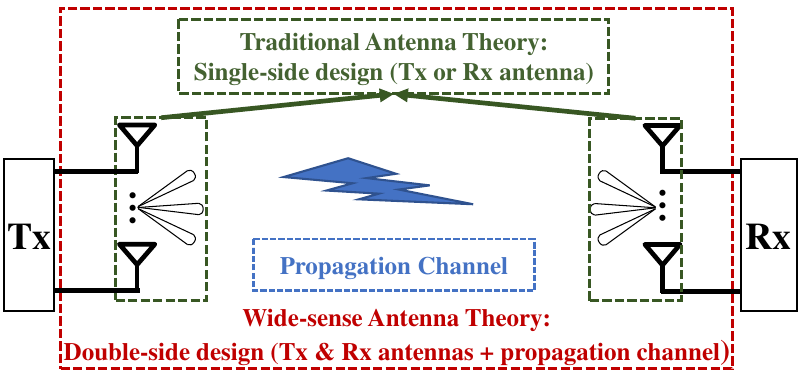}}
	\caption{The description of the traditional antenna theory and wide-sense antenna theory.}
	\label{5}
\end{figure}

As one of the key technologies of wireless communication networks, compact MIMO antenna arrays are capable of improving spectral efficiency. The operational BSs tend to have limited aperture size and the wireless devices are becoming increasingly miniaturized. When the antenna elements are closely packed for having compact construction, strong mutual coupling occurs between adjacent elements, hence potentially resulting in antenna mismatch and antenna pattern distortion. The classic calculation of channel capacity does not consider important antenna parameters such as the antenna pattern, impedance, and mutual coupling coefficient. Additionally, the traditional antenna theory considers the single-sided Tx or Rx antenna design and mainly aims for reducing power loss, but does not consider optimizing the statistical properties and the double-sided Tx and Rx channel capacity\cite{ref_wen}. However, the antennas and the communication environments are inseparable in ultra-massive MIMO schemes, and traditional antenna theory cannot design new antennas by considering single-sided Tx or Rx antenna parameters. It is necessary to propose wide-sense antenna theory that considers both the parameters of the double-sided Tx and Rx antennas, channel properties, and channel capacities. The traditional antenna theory and wide-sense antenna theory are contrasted in Fig.~\ref{5}. By exploiting the angle of arrival and departure, the antenna lobe can be optimized based on the communication environment and can guide the ultra-massive MIMO design. According to the channel's statistical properties such as its space-time-frequency correlation functions, bespoke reconfigurable antennas,  holographic MIMO schemes, and intelligent metasurface unit selection arrangements can be designed. Therefore, the intrinsic integration of information theory and antenna theory holds this promise. 

\section{Conclusions}
In comparison to 5G, the emergence of key 6G technologies brings about novel applications and technical requirements that surpass the application scope of each of the four single theories. In 6G space-air-ground-sea networks, the underlying wireless channels tend to be of continuous nature in the 3D space and the wireless propagation channels are inseparable from the antennas. Therefore, the confluence of electromagnetic theory, information theory, wireless propagation channel modeling theory, and antenna theory is inevitable. Hence, we have characterized the relationships and the gaps among the four fundamental theories. Then, several ways and applications of integrating these four theories into the EIT have been discussed. In summary, EIT research will establish a solid theoretical basis for 6G systems and facilitate significant breakthroughs.

\section*{Acknowledgment}
\small {This work was supported by the National Natural Science Foundation of China (NSFC) under Grants 62394290, 62394291, 61960206006, 62271147 and 62288101, the Fundamental Research Funds for the Central Universities under Grants 2242022k60006 and 2242023K5003, the Key Technologies R$\&$D Program of Jiangsu (Prospective and Key Technologies for Industry) under Grants BE2022067 and BE2022067-1, the EU H2020 RISE TESTBED2 project under Grant 872172, the High Level Innovation and Entrepreneurial Doctor Introduction Program in Jiangsu under Grant JSSCBS20210082, and the Start-up Research Fund of Southeast University under Grant RF1028623029. L. Hanzo would like to acknowledge the financial support of the Engineering and Physical Sciences Research Council projects EP/W016605/1, EP/X01228X/1 and EP/Y026721/1 as well as of the European Research Council's Advanced Fellow Grant QuantCom (Grant No. 789028).
}
\bibliographystyle{IEEEtran}

\section*{Biographies}

\begin{IEEEbiographynophoto}{Cheng-Xiang Wang}
	(Fellow, IEEE) received the Ph.D. degree in wireless communications from Aalborg University, Denmark, in 2004. He is currently a Professor and Executive Dean of the School of Information Science and Engineering, Southeast University, China. He has published more than 540 papers in journals and conference proceedings, including 30 highly cited papers. His current research interests include wireless channel measurements and modeling, 6G wireless communication networks, and electromagnetic information theory. He is a Member of the Academia Europaea, a Member of the European Academy of Sciences and Arts, a Fellow of the Royal Society of Edinburgh and IET.
\end{IEEEbiographynophoto}
\begin{IEEEbiographynophoto}{Yue Yang}
	received the B.E. degree in Communication Engineering from Xidian University, China, in 2020. She is currently pursuing the Ph.D. degree in the National Mobile Communications Research Laboratory, Southeast University, China. Her research interests include electromagnetic information theory, 6G wireless channel modeling and characteristic analysis, and 6G wireless communications.
\end{IEEEbiographynophoto}
\begin{IEEEbiographynophoto}{Jie Huang} 
	(Member, IEEE) received the B.E. degree in Information Engineering from Xidian University, China, in 2013, and the Ph.D. degree in Information and Communication Engineering from Shandong University, China, in 2018. He is currently an Associate Professor in the National Mobile Communications Research Laboratory, Southeast University, China and also a researcher in Purple Mountain Laboratories, China. His research interests include millimeter wave, THz, massive MIMO, reconfigurable intelligent surface channel measurements and modeling, wireless big data, and 6G wireless communications. 
\end{IEEEbiographynophoto}

\begin{IEEEbiographynophoto}{Xiqi Gao}
	(Fellow, IEEE) received the Ph.D. degree in electrical engineering from Southeast University, Nanjing, China, in 1997. He received the Science and Technology Awards of the State Education Ministry of China in 1998, 2006 and 2009, the National Technological Invention Award of China in 2011, the Science and Technology Award of Jiangsu Province of China in 2014, and the 2011 IEEE Communications Society Stephen O. Rice Prize Paper Award in the field of communications theory. His research interests include broadband multicarrier communications, massive MIMO wireless communications, satellite communications, optical wireless communications, information theory and signal processing for wireless communications.
\end{IEEEbiographynophoto}
\begin{IEEEbiographynophoto}{Tie Jun Cui}
	(Fellow, IEEE) received the B.Sc., M.Sc., and Ph.D. degrees in electrical engineering from Xidian University, Xi’an, China, in 1987, 1990, and 1993, respectively. He is the director of the State Key Laboratory of Millimeter Wave, and the Founding Director of the Institute of Electromagnetic Space, Southeast University. Dr. Cui is the author of three books (Springer, 2009; CRC, 2016; Cambridge University Press, 2021), and has published over 600 peer-reviewed journal articles with more than 64 000 citations. He is the editor and editorial board member for several prestigious journals. Dr. Cui is an academician of the Chinese Academy of Science. His research interests include metamaterials and computational electromagnetics. 
\end{IEEEbiographynophoto}
\begin{IEEEbiographynophoto}{Lajos Hanzo}
	(Life Fellow, IEEE) received the M.Sc. and Ph.D. degrees from the Technical University (TU) of Budapest, in 1976 and 1983, respectively. He was also awarded the Doctor of Sciences (DSc) degree by the University of Southampton (2004) and Honorary Doctorates by the TU of Budapest (2009) and by the University of Edinburgh (2015). He is a Foreign Member of the Hungarian Academy of Sciences and a former Editor-in-Chief of the IEEE Press. He has served several terms as Governor of both IEEE ComSoc and of VTS. He is also a Fellow of the Royal Academy of Engineering (FREng), of the IET and of EURASIP. He holds the Eric Sumner Technical Field Award.
\end{IEEEbiographynophoto}
\end{document}